\documentclass{PoS}
\usepackage{multicol}
\title{Charmed Hadron Interactions}

\ShortTitle{Charmed Hadron Interactions}

\author{\speaker{Liuming Liu}\\
        Department of Physics,College of
William and Mary, Williamsburg, VA, 23187, USA \\
Jefferson Laboratory, Newport News, VA, 23606, USA\\
        E-mail: \email{lxliux@wm.edu}}

\author{Huey-Wen Lin\\
Jefferson Laboratory, Newport News, VA, 23606, USA\\
        E-mail: \email{hwlin@jlab.org}}

\author{Kostas Orginos\\
        Department of Physics,College of
William and Mary, Williamsburg, VA, 23187, USA\\
 Jefferson Laboratory,
Newport News, VA, 23606, USA\\
        E-mail: \email{kostas@jlab.org}}

\abstract{We compute the scattering lengths of   charmed mesons and
charmonia scattering  with light hadrons in full QCD. We use Fermilab
formulation for the charm quark and  domain-wall fermions for the
light quarks and staggered sea quarks.  Four different light-quark
masses are used to extrapolate to the physical point. The charmed
baryon spectrum is also presented.}

\FullConference{The XXVI International Symposium on Lattice Field Theory\\
         July 14-19 2008\\
         Williamsburg, Virginia, USA}

\begin{document}

\section{Introduction}

Lattice QCD calculations of  the properties of hadronic interactions
such as elastic scattering phases shifts and scattering lengths have
recently started to develop. Precision results have been obtained in
the  meson-meson sector for certain processes such as pion-pion,
kaon-kaon and pion-kaon scattering and preliminary results for
baryon-baryon scattering lengths have been presented.  A recent
review of these calculations can be found in\cite{Hadronic
interactions}. In this work we study scattering processes where one
hadron contain a charm quark. Firstly, we study the scattering
processes of chamonia ($\eta_c$ and $J/\Psi$) with light hadrons
($\pi, \rho$ and $N$). As it has been pointed out in the
literature~\cite{Brodsky:1989jd,Luke:1992tm,Brodsky:1997gh}, such
interaction has  a direct relation to possible charmonium-nucleus
bound states with binding energy of a few MeV. Unlike the
traditional nuclear force that binds nucleons, in this case, there
are no quark exchange diagrams, and only gluons  are responsible for
the binding. In other words, the charmonium nucleon force is purely
a gluonic  van der Waals force. The charmonium interactions with
light  hadrons has been also studied  in quenched lattice
QCD~\cite{Charmonium hadron}. Secondly, we study the scattering
processes of charmed mesons ($D_s $ and $D$) with light mesons
($\pi$ and $K$). In reference \cite{Dpi Dkaon}, $D\pi$ and $DK$
scattering were studied using the scalar form factors in
semileptonic pseudoscalar-to-pseudoscalar decays. In our work we use
L\"{u}scher's formula \cite{Two body} to extract scattering
information from the energy shift of two interacting hadrons
relative to the total energy of the two individual hadrons.
\section{Fermion Actions}
We use Fermilab formulation \cite{FermiLab formulation} for the
charm quark, domain-wall fermions for the light quarks and staggered
sea quarks. In the following, we will specify the Fermilab
formulation as well as the tuning of the parameters in this
formulation. The action is:
\begin{eqnarray*}
  S &=& S_0+S_B+S_E , \\
  \label{eq:action_S_0} S_0 &=& \sum_x \bar{q}(x)[m_0+(\gamma_0
  \nabla_0-\frac{b}{2}\triangle_0) + \nu
  \sum_i(\gamma_i\nabla_i-\frac{b}{2}\triangle_i)]q(x) , \\
  \label{eq:action_S_B} S_B &=& -\frac{b}{2}c_B\sum_x
  \bar{q}(x)(\sum_{i<j}\sigma_{ij}F_{ij})q(x) ,\\
  \label{eq:action_S_E} S_E &=& -\frac{b}{2}c_E\sum_x
  \bar{q}(x)(\sum_i\sigma_{0i}F_{0i})q(x),
\end{eqnarray*}
  where $b$ is the lattice spacing, $\nabla_0$ and $\nabla_i$
  are first-order lattice derivatives in time direction and space directions, $\triangle_0$
  and $\triangle_i$ are second-order lattice derivatives, $F_{\mu\nu}$ is the field tensor defined in reference \cite{FermiLab formulation}.
  In this action, the space-time exchange symmetry is not imposed.
  $S_0$ is just the standard Wilson fermion action except that the
  coefficient in front of the space term and the coefficient in
  front of the temporal term are different. $S_B$ and $S_E$ are
  spatial and temporal clover terms, also with different
  coefficients.\\
  \indent There are four parameters to tune: the charm quark mass $m_c$, the
  anisotropy $\nu$, and the two clover coefficients $c_B$ and $ c_E$.
  We use the spin-average mass of $J/\Psi$ and $\eta_c$ to tune the
  charm quark mass. The value of $\nu$ is tuned to restore the
  dispersion relation. As for $c_B$ and $c_E$, the tree level tadpole
  improvement estimate is $c_B=c_E=1/u_0^3$. Chen suggested a better
  way to evaluate the two parameters \cite{Clover coefficients}:
  \begin{equation}
           c_B=\frac{\nu}{u_0^3}, \quad \quad
           c_E=\frac{1}{2}(1+\nu)\frac{1}{u_0^3} \nonumber .
  \end{equation}
  Here $c_B$ and $c_E$ both depend on $\nu$.  We use Chen's
  evaluation in our work.
  \section{Numerical Ensembles}
  We employ the gauge configurations generated by the MILC
  Collaboration \cite{Configuration}. We use the $20^3\times64$ lattices generated at four
  values of light-quark masses. The lattice spacing $b=0.12406$ fm.
  The details of the ensembles are listed below:
  \begin{center}
  \begin{tabular}{c|ccccc}

  \hline\hline
  Ensemble             &$bm_l$      &$bm_s$     & $bm_l^{dwf}$   & $bm_s^{dwf}$ & number of props \\
  \hline
  2064f21b676m007m050  &0.007       &0.050      & 0.0081         & 0.081        & 450\\
  2064f21b676m010m050  &0.010       &0.050      & 0.0138       &0.081         & 650\\
  2064f21b679m020m050  &0.020       &0.050      & 0.0313      & 0.081        & 550\\
  2064f21b781m030m050  &0.030       &0.050       & 0.0478      & 0.081         & 380\\
  \hline\hline

  \end{tabular}
  \end{center}
  The subscript $l$
  denotes light quark, and $s$ denotes the strange quark. The superscript $dwf$ denotes domain-wall fermion.

  \section{Heavy-Quark Action Test}
  In the heavy-quark action, the dispersion relation $E^2=m^2+c^2p^2$ needs to
  be restored by tuning the parameter $\nu$ to get the value of $c^2$ to agree with the theoretical value
  $1$. To do that, we calculate the single-particle energy of $\eta_c$,
  $J/\Psi$, $D_s$ and $D$ at the six lowest momenta: $\frac{2\pi}{L}(0,0,0),
   \frac{2\pi}{L}(1,0,0),$ $ \frac{2\pi}{L}(1,1,0),\frac{2\pi}{L}(1,1,1), \frac{2\pi}{L}(2,0,0),
  \frac{2\pi}{L}(2,1,0)$. We tune the dispersion relation of $\eta_c$ and get the dispersion relations of
  $J/\Psi$, $D_s$ and $D$ to be restored as well. The following table lists the values of
  $c^2$ we get from the numerical simulation:
       \begin{center}
       \begin{tabular}{|c|c|c|c|c|}
       \hline
       $ $           &$\eta_c$         &$J/\Psi$            &$D$              &$D_s$              \\
       \hline\hline
       $c^2$       &$0.989(0.005)$   &$0.965(0.009)$    &$1.012(0.017)$    &$1.006(0.009)$\\
       \hline
       \end{tabular}
       \end{center}
By tuning the charm-quark mass, we get the spin-average mass of $\eta_c$ and $J/\Psi$ to be $3056.54(1.15)\mathrm{MeV}$, which agree well
  with the experimental value. We also calculate the hyperfine
  splitting, which we find to be $99.1(1.1)\mathrm{Mev}$.

  In this section, all numerical results are obtained on the ensemble
  2064f21b676m007m050 which has the lightest light-quark mass. We keep the charm
  quark mass and anisotropy parameter fixed for all ensembles.

  \section{Baryon Spectrum}
\begin{figure}
   \vspace{-3in}
  \begin{multicols}{4}
  \hspace{-1cm}
  \includegraphics[width=1\textwidth]{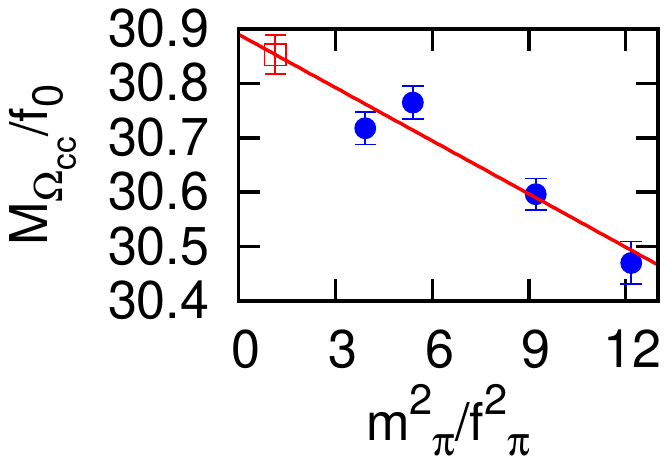}

  \hspace{-1cm}
   \includegraphics[width=1\textwidth]{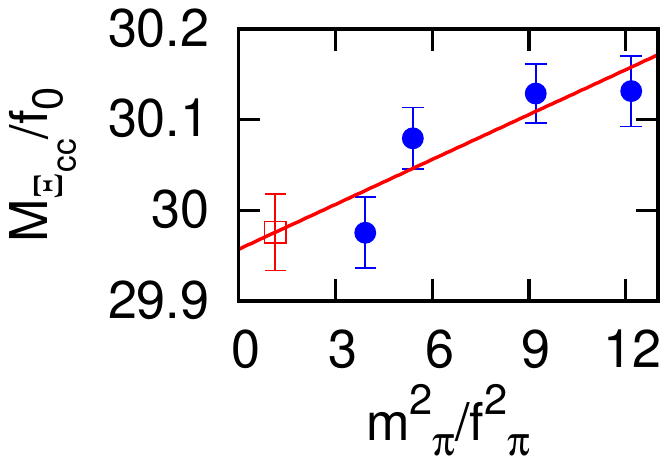}

   \hspace{-1cm}
    \includegraphics[width=1\textwidth]{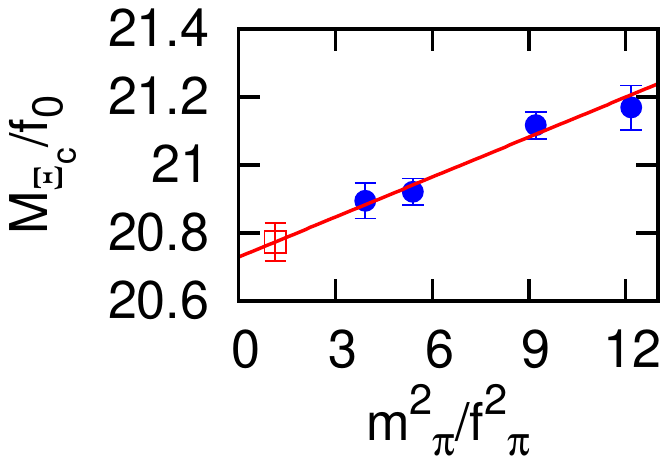}

    \hspace{-1cm}
     \includegraphics[width=1\textwidth]{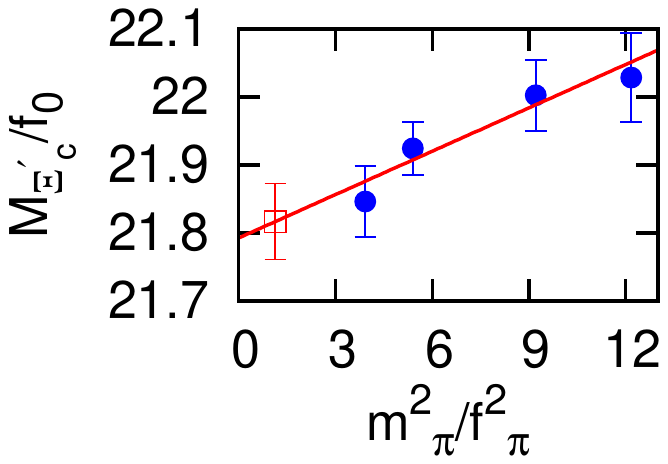}
  \end{multicols}

   \vspace{-3.5in}
  \begin{multicols}{4}
  \hspace{-1cm}
  \includegraphics[width=1\textwidth]{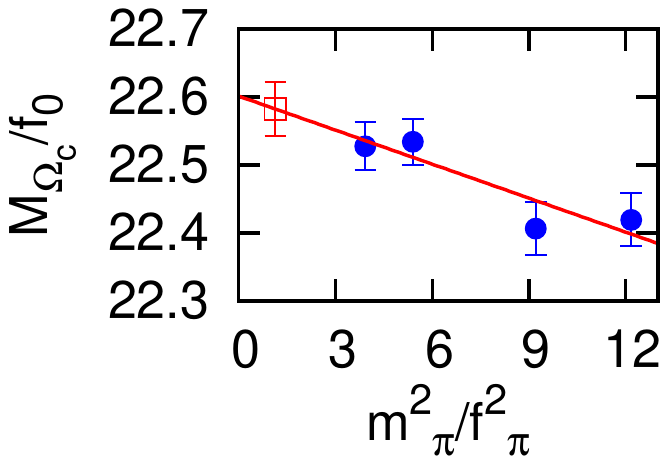}

  \hspace{-1cm}
   \includegraphics[width=1\textwidth]{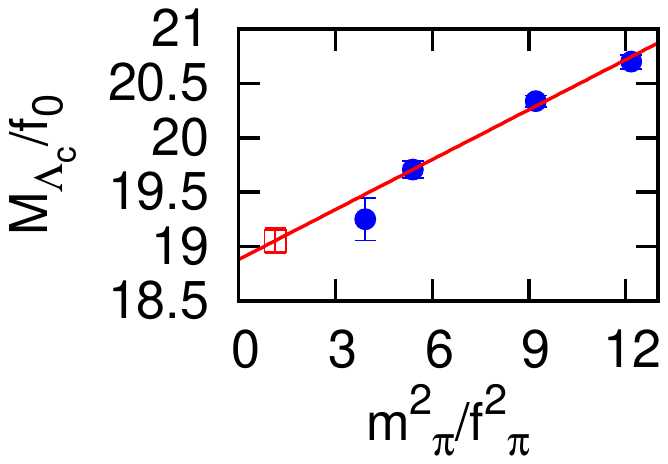}

   \hspace{-1cm}
    \includegraphics[width=1\textwidth]{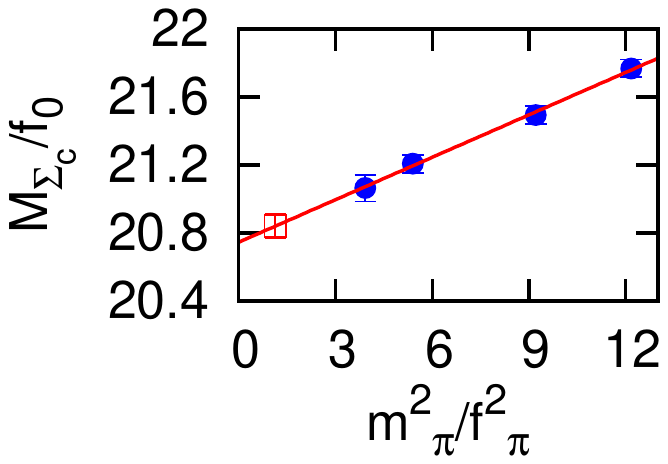}
   \end{multicols}

 \vspace{-0.5in}
  \caption{Charmed baryon masses as functions of $m_\pi^2/f_\pi^2$. The masses are divided by $f_0=130.7\mathrm{MeV}$ to make them dimensionless.
  The four blue points in each
  panel denote the baryon masses at four different light-quark masses. They are extrapolated to the
  physical values, which are denoted by
  the red points.  }
 \label{fig:baryon}
 \end{figure}

\begin{figure}
  \vspace{-1in}
  \hspace{1in}
  \includegraphics[width=16cm]{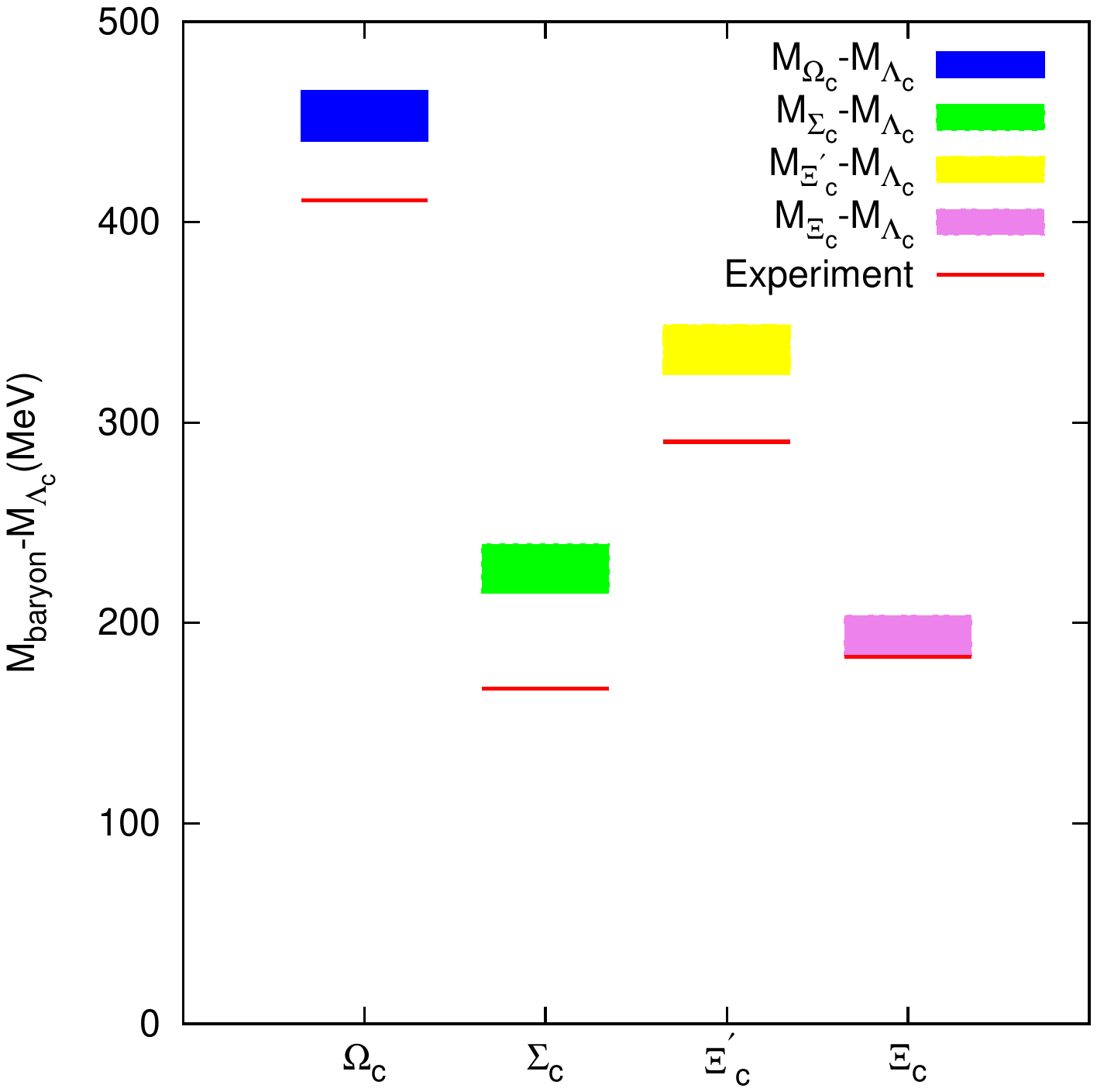}
  \caption{Mass splittings of singly charmed baryons. The red lines represent the experimental values.
  The four bars in different colors
  represent the numerical values of the mass difference between $\Lambda_c$ and other singly charmed baryons.
  The heights of the bars
  indicate the statistical errors.}
 \label{fig:splitting}
 \end{figure}

  We calculate the masses of singly charmed baryons and doubly charmed
  baryons at four different light quark masses. We employ a simple linear
  relation $m_{baryon}=c_1 + c_2 m_{\pi}^2/f_{\pi}^2$ to extrapolate
  the baryon masses to the physical point. The values of $m_{\pi}/f_{\pi}$ are taken from
  reference \cite{m_pi/f_pi}.
  Fig. \ref{fig:baryon} shows
  the baryon mass spectrum. We present the singly charmed baryon mass
  splittings in Fig. \ref{fig:splitting}. Our results are comparable with similar work  done using a staggered light-quark  a
  ction \cite{Baryon spectrum}.

  \section{Charmed Hadron Interactions}

\begin{figure}
  \vspace{-3.5in}
   \begin{multicols}{3}
  \includegraphics[width=1.2\textwidth]{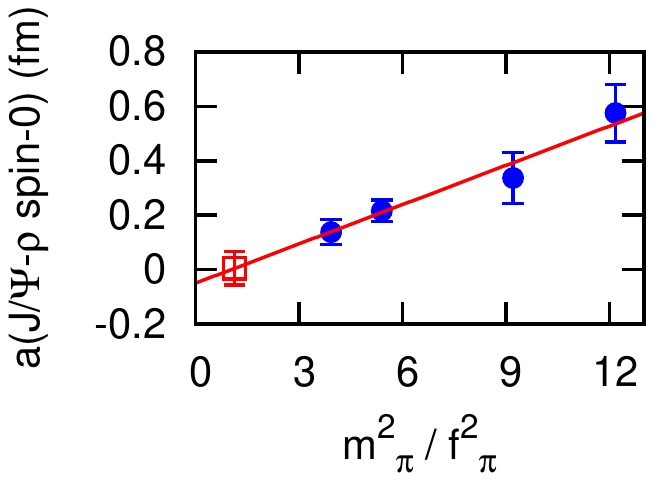}
   \includegraphics[width=1.2\textwidth]{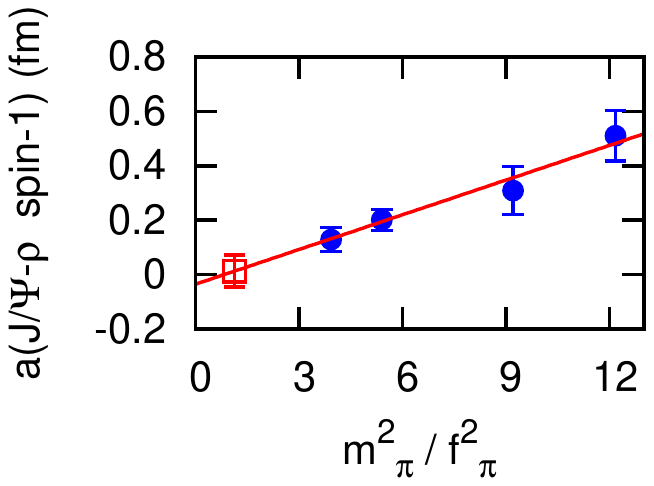}
    \includegraphics[width=1.2\textwidth]{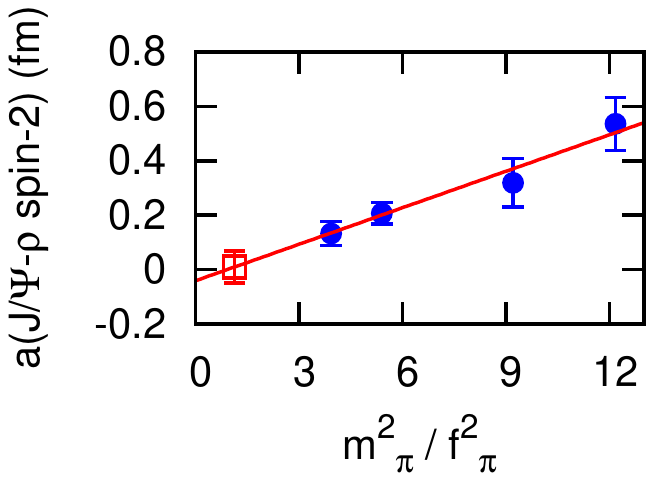}
  \end{multicols}

  \vspace{-4in}
     \begin{multicols}{3}
  \includegraphics[width=1.2\textwidth]{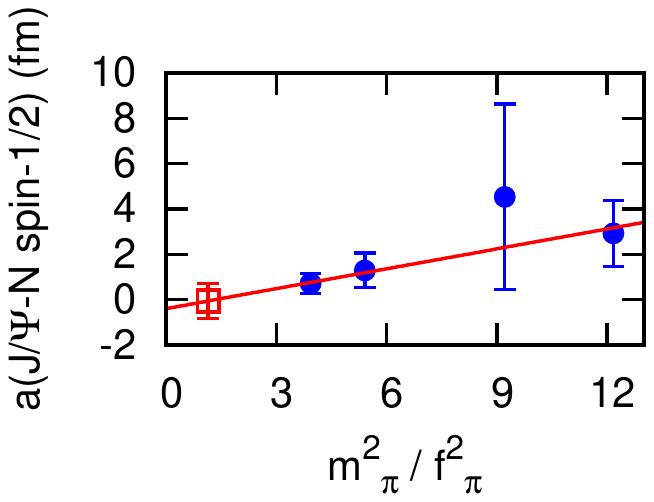}
   \includegraphics[width=1.2\textwidth]{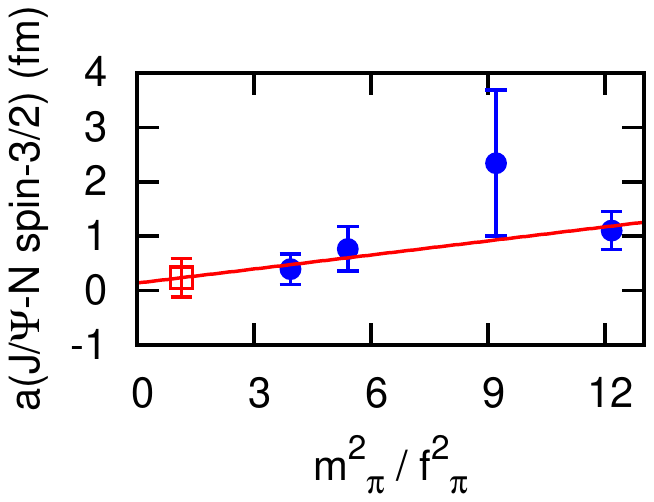}
    \includegraphics[width=1.2\textwidth]{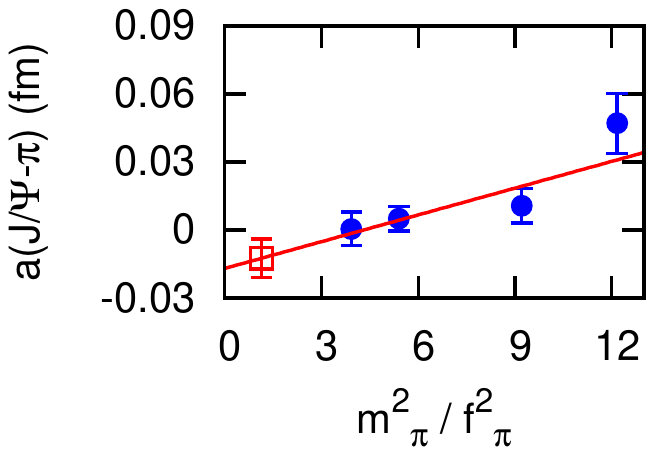}
  \end{multicols}

  \vspace{-4in}
   \begin{multicols}{3}
  \includegraphics[width=1.2\textwidth]{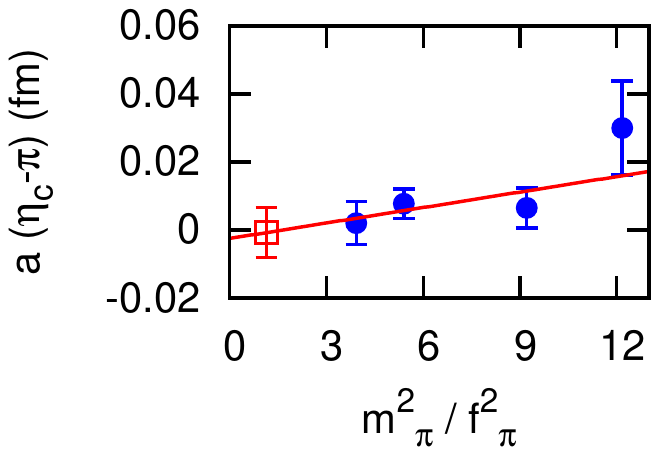}
   \includegraphics[width=1.2\textwidth]{ep.pdf}
    \includegraphics[width=1.2\textwidth]{ep.pdf}
  \end{multicols}

  \vspace{-0.5in}
  \caption{ Charmonium to pion, rho and nucleon scattering lengths as functions of $m_\pi^2/f_\pi^2$. In each panel, the four blue points
  denote the scattering length
  measured at four different light-quark masses. The red points denote the extrapolated values. }
  \label{fig:slen1}
  \end{figure}

\begin{figure}
  \vspace{-2.7in}
   \begin{multicols}{4}
   \hspace{-1cm}
  \includegraphics[width=1\textwidth]{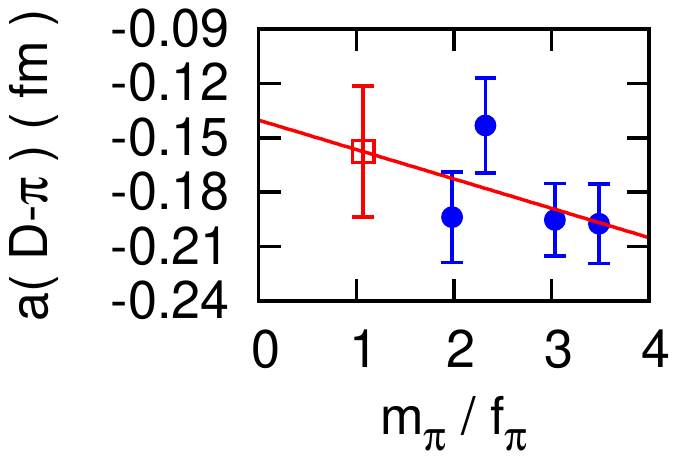}

  \hspace{-1cm}
   \includegraphics[width=1\textwidth]{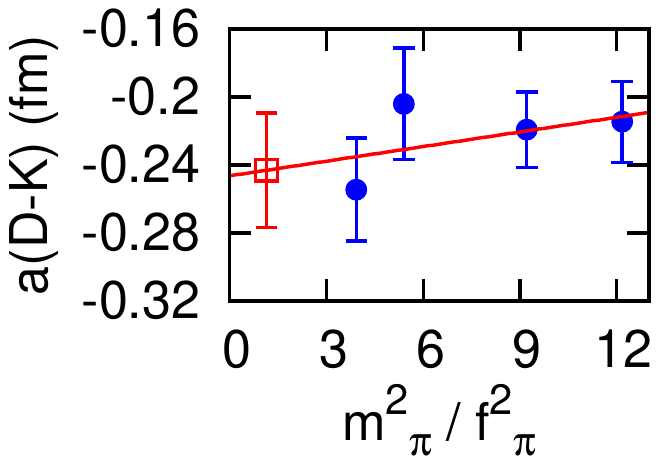}

   \hspace{-1cm}
    \includegraphics[width=1\textwidth]{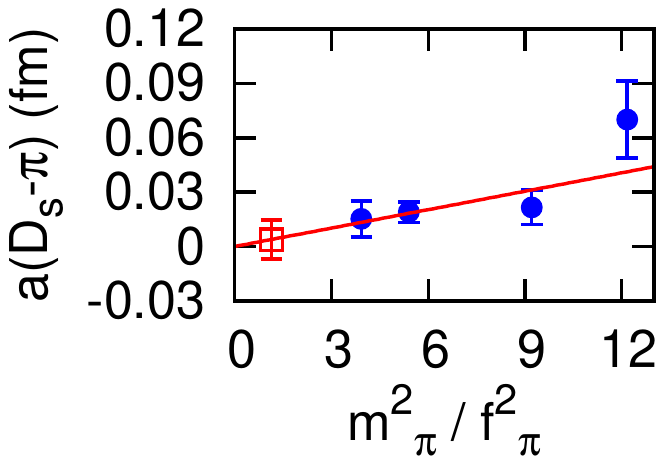}

    \hspace{-1cm}
     \includegraphics[width=1\textwidth]{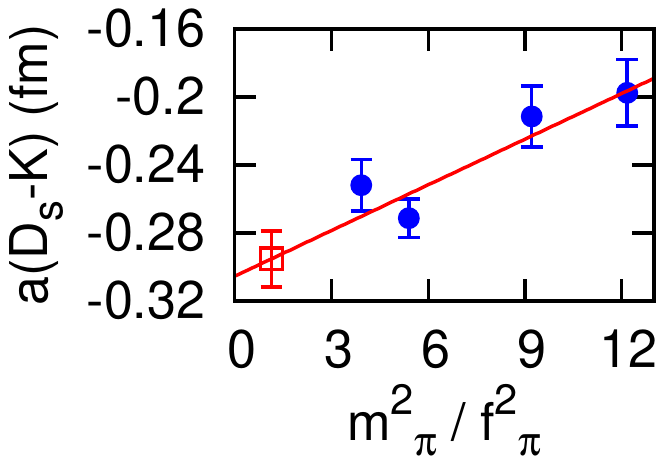}
  \end{multicols}

  \vspace{-0.5in}
  \caption{ Charmed meson scattering lengths as functions of $m_\pi^2/f_\pi^2(m_\pi/f_\pi)$. In each panel, the four blue points
  denote the scattering length
  measured at four different light-quark masses. The red points denote the extrapolated values. }
  \label{fig:slen2}
  \end{figure}

 L\"{u}scher has shown
that the scattering phase shift is related to the energy shift
($\triangle E$) of two interacting hadrons relative to the total
energy of the two individual hadrons \cite{Two body}.  The total
energy of two interacting hadrons ($h_1$ and $h_2$) is obtained from
the four-point correlation function:
\begin{equation}
G^{h_1-h_2}(t)=\langle \mathcal{O}^{h_1}(t)
\mathcal{O}^{h_2}(t)(\mathcal{O}^{h_1}(0) \mathcal{O}^{h_2}(0))^\dag
\rangle .
\end{equation}
To extract the energy shift $\triangle E$, we define a ratio
$R^{h_1-h_2}(t)$:
\begin{equation}
R^{h_1-h_2}(t)=\frac{G^{h_1-h_2}(t,0)}{G^{h_1}(t,0)G^{h_2}(t,0)}\longrightarrow
\exp(-\triangle E \cdot t) ,
\end{equation}
where $G^{h_1}(t,0)$ and $G^{h_2}(t,0)$ are two-point functions.
$\triangle E$ can be obtained by fitting $R^{h_1-h_2}(t)$ to a
single exponential.
\\
\indent The magnitude of center-of-mass momentum $p$ is related to
$\triangle E$ by
\begin{equation}
\label{equ:deltaE}
 \triangle E = \sqrt{p^2+m_{h_1}^2} +
\sqrt{p^2+m_{h_2}^2} - m_{h_1} - m_{h_2} .
\end{equation}
The phase shift is obtained from the following relation:
\begin{equation}
 \label{equ:pcotdelta1}
p\cot \delta(p) = \frac{1}{\pi
L}\textbf{S}\Big(\Big(\frac{pL}{2\pi}\Big)^2\Big) ,
\end{equation}
  where the $\textbf{S}$ function is defined as
  \begin{equation}
  \textbf{S}(x)=\sum_{\textbf{j}}^{|\textbf{j}| < \Lambda}
  \frac{1}{|\textbf{j}|^2-x}-4\pi\Lambda .
  \end{equation}
  The sum is over all three-vectors of integers $\textbf{j}$ such
  that $|\textbf{j}|< \Lambda$, and the limit $\Lambda \rightarrow \infty$ is implicit.
  If the interaction range
  is smaller than half of the lattice size, the $s$-wave phase shift
  will be
  \begin{equation}
  \label{equ:pcotdelta2}
  p\cot \delta(p) = \frac{1}{a} + \mathcal {O}
  (p^2) ,
  \end{equation}
  where $a$ is the scattering length. \\
  \indent By measuring the energy shift, the momentum $p$ can be obtained by equation (\ref{equ:deltaE}).
  Then we calculate the right-hand side of equation (\ref{equ:pcotdelta1}). From equation (\ref{equ:pcotdelta1}) and
  equation (\ref{equ:pcotdelta2}), we can see that the scattering length is just the inverse of the right-hand side
  of equation (\ref{equ:pcotdelta1}). \\
  \indent We study the scattering of the charmonia ($\eta_c,
  J/\Psi$) with light hadrons ($\pi$, $\rho$, N). The scattering of charmed mesons ($D_s, D$)
  with $\pi$ and $K$ are also studied. We choose the isospin-$\frac{3}{2}$ channel for $D-\pi$ scattering
  and isospin-$1$ channel for $D-K$ scattering to avoid disconnected diagrams.
  For the same reason, we use $D_s^+$ and $K^+$ for $D_s-K$ scattering.
  The interactions of the heavy hadron($h$) with the pion has a
  special feature due to the Nambu-Goldstone nature of the pion. The
  $s$-wave scattering length is given by~\cite{Weinberg:1966zz}
  \begin{equation}
  \label{equ:slen}
  a^{h-\pi}=-(1+\frac{m_\pi}{m_h})^{-1}\frac{m_\pi}{8\pi
  f_\pi^2}[I(I+1)-I_h(I_h+1)-2]+\mathcal{O}(m_\pi^2),
  \end{equation}
  where I is the total isospin of the $\pi-h$ system. The first term
  in equation (\ref{equ:slen}) vanishes for $J/\Psi-\pi$, $\eta_c-\pi$ and $D_s-\pi$
  channels but not for $D-\pi$ channel. So the leading term of $D-\pi$ scattering length
  is proportional to $m_\pi/f_\pi$ while the leading terms of $J/\Psi-\pi$, $\eta_c-\pi$ and $D_s-\pi$ scattering lengths are proportional
  to $m_\pi^2/f_{\pi}^2$ . We measure the scattering
  lengths at four different light-quark masses. We use the relation $a=c_1+c_2
  m_\pi/f_{\pi}$ to extrapolate $D-\pi$ scattering length to the
  physical point. For other channels, we use the relation $a=c_1+c_2
  m_\pi^2/f_{\pi}^2$ to do the extrapolations because the leading term in equation (\ref{equ:slen}) vanishes.
  Fig. \ref{fig:slen1} and Fig. \ref{fig:slen2} show the scattering lengths of
all these
  channels.


\newpage
  The following tables list the extrapolated scattering lengths of all channels:

\begin{multicols}{2}

  \begin{tabular}{|c|c|}
       \hline
        Channel                 & Scattering lengths(fm)                 \\
       \hline\hline
       $\eta_c-\pi$            &$0.00(1)$                                  \\
       \hline
       $\eta_c-\rho$            &$0.03(5)$                             \\
       \hline
       $\eta_c-N$                 &$0.18(9)$                               \\
       \hline
       $J/\Psi-\pi$               &$-0.01(1)$                               \\
       \hline
       $J/\Psi-N \, $ spin $1/2$            &$-0.05(77)$            \\
       \hline
       $J/\Psi-N \, $ spin  $3/2$            &$0.24(35)$           \\
       \hline
       $J/\Psi-\rho \,$ spin $0$                 &$0.00(6)$   \\
       \hline
          \end{tabular}

   \begin{tabular}{|c|c|}
       \hline
        Channel                 & Scattering lengths(fm)             \\
       \hline\hline
          $J/\Psi-\rho \,$ spin $1$               &$0.01(6)$ \\
       \hline
     $J/\Psi-\rho \,$ spin $2$               &$0.01(6)$\\
       \hline
     $D-\pi$                                 &$-0.16(4)$\\
       \hline
    $D-K$                                     &$-0.23(4)$\\
       \hline
     $D_s-\pi$                                  &$0.00(1)$  \\
       \hline
      $D_s-K$                                  &$-0.31(2)$ \\
       \hline
             \end{tabular}
          \end{multicols}

  \section{Conclusion}
  We calculated the scattering lengths of charmed mesons and charmonia with light hadrons.
  For the channels of charmonia with light hadrons and
  $D_s-\pi$ channel, we found weak interactions. The scattering lengths are zero or close to
  zero. For the $D-\pi, D-K$ and $D_s-K$ channels, we found relatively
  strong repulsive interactions. In the future, we will improve statistics to get more reliable  results,
  and perform extrapolations using chiral perturbation theory forms.

\section{Acknowledgement}
We would like to thank USQCD for the computing resources used to carry  out
this study, and NPLQCD collaboration for providing the light quark propagators. Additional  analysis was done on the {\it cyclades}
cluster at WM. This work was partially supported by the US Department of Energy, under contract nos. DE-AC05-06OR23177(JSA),
DE-FG02-07ER41527, and DE-FG02-04ER41302; and by the Jeffress Memorial Trust, grant J-813.

\end{document}